\documentclass[11pt]{article}

\usepackage{graphicx}
\usepackage{amsmath}
\usepackage{amsfonts}
\usepackage{fancyhdr} 

\usepackage{wrapfig}

\usepackage{bbold}

\newtheorem{lem}{Lemma}
\newtheorem{thm}{Theorem}

\newtheorem{prop}[lem]{Proposition}
\newtheorem{cor}[lem]{Corollary}
\newtheorem{defn}[lem]{Definition}
\newtheorem{alg}{Algorithm}

\numberwithin{equation}{section}
\numberwithin{thm}{section}
\numberwithin{lem}{section}
\numberwithin{alg}{section}

\newcommand{\real}{\mathbb{R}}
\newcommand{\integer}{\mathbb{Z}}

\newcommand{\ddd}{\mathbb{D}}

\newcommand{\complex}{\mathbb{C}}

\newenvironment{pf}{\noindent {\em Proof}.\ \ }{\hspace*{\fill}\rule{.5ex}{1.4ex}\,}

\newcommand{\lll}{\mathcal L}
\newcommand{\one}{\mathbb{1}}
\newcommand{\hhh}{\mathcal{H}}
\newcommand{\ggg}{\mathcal{G}}
\newcommand{\ccc}{\mathcal{C}}
\newcommand{\wcc}{\widehat{\mathcal{C}}}
\newcommand{\ttt}{\mathbb{T}}

\title{\vspace*{-1in}Hardy space on the polydisk and scattering in layered media}
\author{Peter C~Gibson\footnote{Dept.~of Mathematics \& Statistics, York University, 4700 Keele St., Toronto, Ontario, Canada, M3J~1P3, $\mathtt{pcgibson@yorku.ca}$} }

\begin{document}

\maketitle

\begin{abstract} 
Hardy space on the polydisk provides the setting for a global description of scattering in piecewise-constant layered media, giving a simple qualitative interpretation for the nonlinear dependence of the {G}reen's function on reflection coefficients and layer depths.   Using explicit formulas for amplitudes, we prove that the power spectrum of the {G}reen's function is approximately constant.  In addition we exploit a connection to Jacobi polynomials to derive formulas for computing reflection coefficients from partial amplitude data.   Unlike most approaches to layered media, which variously involve scaling limits, approximations or iterative methods, the formulas and methods in the present paper are exact and direct.

 \end{abstract}

\newpage
\tableofcontents
\newpage

\section{Introduction}

The theory of wave propagation in layered media is important for various imaging modalities, including acoustic, seismic and electromagnetic imaging.  For example, in geophysics a layered half space serves as a simple model for stratified rock layers that are characteristic of sedimentary formations targeted in oil exploration.  From the mathematical perspective, the equation governing wave propagation in a piecewise constant layered half space serves as a basic example of a PDE whose coefficients are discontinuous and which therefore falls outside the scope of a good deal of established theory  (such as Hormander-Duistermaat theory of FIOs, Gelfand-Levitan methods, and so on).   The present paper is principally concerned with the reflection Green's function for such a PDE, and the nonlinear dependence of the Green's function on physical  parameters.  From the imaging perspective, the reflection Green's function roughly corresponds to measured data, and the essential inverse problem is to infer physical structure from the data.   

Despite the considerable body of existing literature (see \cite{FoGaPaSo:2007}, \cite{BrGo:1990} and the many references therein), there are some basic theoretical questions that have yet to be answered, including the following.
\begin{enumerate}
\item  How precisely do Green's function amplitudes depend on reflection coefficients?
\item Numerical experiments show that the power spectrum of the Green's function tends to be approximately constant---why is this?
\item Is it possible to determine reflection coefficients using localized amplitude data?
\end{enumerate}

The purpose of the present paper is to introduce a new deterministic perspective on piecewise constant layered media, and to exploit this perspective to answer the questions above.  In the process we establish connections to orthogonal polynomials, holomorphic functions on the polydisk, and almost periodic functions, all of which supply machinery applicable to the problems at hand.   Further to the three questions, an additional issue motivated the present paper.   Recent investigations into minimum phase preserving operators \cite{GiLa:2012} give indirect evidence that Hardy space should somehow be connected to PDEs modeling the propagation of seismic waves---but without showing how.  The present paper clarifies this issue by providing a direct link.   
 
A crucial first step underlying the solution to (1.) is a combinatorial analysis of scattering sequences, completed recently in \cite{Gi:Comb2013}.   The latter work derives explicit formulas for Green's function amplitudes; and the analysis of these formulas, which is carried out in Section~\ref{sec-amplitudes}, reveals an unexpected connection to Jacobi polynomials.  As a conceptual device, the amplitude formulas may all be combined into a single function on euclidean space, called the covering amplitude (Section~\ref{sec-covering}).  

With the formulas for amplitudes in hand, the next step is to represent the Green's function in terms of polydisk functions (Section~\ref{sec-polydisk}).   Here the notion of an inner function plays a central role, as does the connection between polydisk functions and almost periodic functions---which are obtained as restrictions of polydisk functions to a line on the torus.  This provides a quantitative answer to question (2.), detailed in Section~\ref{sec-power}.    

The polydisk representation illuminates in simple qualitative terms the dependence of the Green's function and its Fourier transform on both reflection coefficients on one hand, and on layer depths on the other hand.  More precisely, amplitudes, which are Taylor coefficients of a polydisk function, can be viewed as the value of the covering amplitude sampled on the twice integer lattice translated by the vector of reflection coefficients.   The Fourier transform of the Green's function is the restriction of the polydisk function to a line on the torus, the sequence of layer depths being the direction vector of the line.   Thus reflection coefficients comprise a translation to be applied to the twice integer lattice, while layer depths determine a line on the torus.   

Turning to the time domain, the support of the Green's function is the image of part of the integer lattice under the action of a particular linear functional, namely scalar multiplication with the vector of layer depths.    We treat the inverse problem of how to recover this linear functional from its values on part of the integer lattice in Section~\ref{sec-arrival}.   This serves as a stepping stone to question (3.), treated in Section~\ref{sec-localized}.   Drawing on the theory of orthogonal polynomials, we show that indeed there is a way to compute reflection coefficients exactly, using only local amplitude data.  We derive an explicit formula, which demonstrates in concrete terms how to exploit the inherent redundancy of reflection data.   

The main object of interest in the present paper is the reflection Green's function.  We also analyze the transmission Green's function, which plays a useful auxiliary role because of conservation of energy (see Section~\ref{sec-conservation}, below).   Our results are theoretical; implementation and testing on experimental data is deferred to a separate paper.

\pagestyle{fancyplain}

\subsection{Preliminaries}

We summarize briefly some standard facts, a detailed derivation of which can be found in \cite[Chapter~3]{FoGaPaSo:2007} (among many other references).  Let $(x,y,z)$ be euclidean coordinates for a three-dimensional solid medium in which the density $\rho$ and bulk modulus $K$ are functions of $z$ alone, referred to as depth.   Suppose further that $\rho$ and $K$ are piecewise constant in $z$, having jumps at the $n+1$ locations 
\[
z_0<z_1<\cdots<z_n
\]
and let $z_{-1}<z_0$ and $z_{n+1}>z_n$ be reference depths in the respective homogeneous half spaces $z<z_0$ and $z>z_n$.   For $0\leq j\leq n+1$, let $K_j$ denote the constant value of the bulk modulus in the layer 
\[
z_{j-1}<z<z_j,
\]
and let $\rho_j$ denote the density in the same layer.  Given initial conditions that depend on $z$ only, the particle velocity $u(t,z)$ and pressure $p(t,z)$ evolve in time $t$ according to the coupled first order equations 
\begin{subequations}
\begin{align}
\rho\frac{\partial u}{\partial t}+\frac{\partial p}{\partial z}&=0\label{wave1}\\
\frac{1}{K}\frac{\partial p}{\partial t}+\frac{\partial u}{\partial z}&=0.\label{wave2}
\end{align}
\end{subequations}
For the sake of definiteness we focus on the velocity field $u(t,z)$, although the results can just as easily be formulated in terms of $p(t,z)$.
The initial conditions corresponding to a plane wave unit impulse propagating from depth $z_{-1}$ are
\begin{equation}\label{initial}
\begin{split}
u(0,z)&=\delta(z-z_{-1})\\
 p(0,z)&=\sqrt{K(z_{-1})\rho(z_{-1})}\;\delta(z-z_{-1}).
 \end{split}
\end{equation}
For $t>0$ this system has a unique solution, $u(t,z)$.   Its restriction to depths $z=z_{-1}$ and $z=z_{n+1}$ are the reflection and transmission Green's functions, respectively,
\begin{equation}\label{Green's}
G(t)=u(t,z_{-1})\quad\mbox{and}\quad H(t)=u(t,z_{n+1}).
\end{equation}
For $0\leq j\leq n+1$, the time it takes a traveling plane wave to go from $z_{j-1}$ to $z_j$ and back is
\begin{equation}\label{tau-formula}
\tau_j=\frac{2(z_j-z_{j-1})}{\sqrt{K_j/\rho_j}}.
\end{equation}
For $0\leq j\leq n$, the reflection coefficient for the interface at depth $z_j$ is 
\begin{equation}\label{R-formula}
R_j=\frac{\sqrt{K_j\rho_j}-\sqrt{K_{j+1}\rho_{j+1}}}{\sqrt{K_j\rho_j}+\sqrt{K_{j+1}\rho_{j+1}}}.
\end{equation}
Write
\[
R=(R_0,\ldots,R_n),\quad\tau=(\tau_0,\ldots,\tau_n)\quad\mbox{and}\quad\tau^\prime=(\tau_0,\ldots,\tau_n,\tau_{n+1}).
\]
The Green's function $G$ is completely determined by the pair $(\tau,R)$, and $H$ by the pair $(\tau^\prime,R)$.   We incorporate this determinacy into the notation, writing 
\[
G^{(\tau,R)}\quad\mbox{and}\quad H^{(\tau^\prime,R)}
\]
for the reflection and transmission Green's functions.  Thus media having a common pair $(\tau^\prime,R)$ of travel times and reflection coefficients are indistinguishable from one another with respect to reflection of waves at the depth $z_{-1}$ or transmission of waves from $z_{-1}$ to $z_{n+1}$.  We shall regard them as the same, and refer to a pair $(\tau^\prime,R)$ or a pair $(\tau,R)$ as a medium, letting it be understood that an equivalence class of media is thereby represented.

\subsection{The backward recurrence\label{sec-backward}}
The standard representation of the Fourier transform of $G^{(\tau,R)}$ is as follows.  Note that to be consistent with \cite{FoGaPaSo:2007}, we use the ``physicist's Fourier transform",
\[
\widehat{f}(\omega)=\int_{-\infty}^\infty f(t)e^{i\omega t}\,dt.
\]
The isometries of the open unit disk $\ddd$ with respect to the Poincar\'e metric
\[
ds^2=\frac{dx^2+dy^2}{(1-x^2-y^2)^2}
\]
are precisely the disk automorphisms $\Psi_\alpha^x:\ddd\rightarrow\ddd$ given by
\begin{equation}\label{automorphism}
\Psi_\alpha^x(z)=e^{i\alpha}\frac{z+x}{1+\bar{x}z}\quad (\alpha\in\real,\;x\in\ddd).
\end{equation}
(See \cite[Chapter~2]{Mi:2006}.)  The Fourier transform of $G^{(\tau,R)}$ is a composition of disk automorphisms indexed by $x=R_j$ and evaluated at 0,
\begin{equation}\label{G-hat}
\widehat{G^{(\tau,R)}}(\omega)=\Psi^{R_0}_{\tau_0\omega}\circ\cdots\circ\Psi^{R_n}_{\tau_n\omega}(0).
\end{equation}
This is referred to as a backward recurrence, since to compute it one has to start with $\Psi^{R_n}_{\tau_n\omega}(0)$ and then recursively evaluate disk automorphisms of decreasing index on the result.  An immediate consequence of this representation is that
\begin{equation}\label{less-than-one}
\left| \widehat{G^{(\tau,R)}}(\omega) \right|\leq 1.
\end{equation} 

\subsection{Conservation of energy\label{sec-conservation}}
In the time domain $G^{(\tau,R)}$ and $H^{(\tau^\prime,R)}$ are delta trains, respectively of the general form
\begin{equation}\label{train}
\begin{split}
G^{(\tau,R)}(t)=&\sum_{j=1}^\infty a_j\,\delta(t-\sigma_j),\\
H^{(\tau^\prime,R)}(t)=&\sum_{j=1}^\infty b_j\,\delta(t-\sigma^\prime_j).
\end{split}
\end{equation}
The coefficients $a_j,b_j$ will be referred to as amplitudes; $\sigma_j$ and $\sigma^\prime_j$ will be called arrival times.   We note a simple but useful consequence of conservation of energy that applies to the amplitudes provided that the arrival times $\sigma_j$ are all distinct and the $\sigma^\prime_j$ are distinct too.   Under these conditions, 
\begin{equation}\label{conservation}
\sum_{j=1}^\infty a_j^2+\sum_{j=1}^\infty b_j^2=1.    
\end{equation}
This expresses the fact that all the energy in the initial pulse (\ref{initial}) is eventually (after scattering within the layers) either reflected back into the half space $z<z_0$ or transmitted into the half space $z>z_n$, whereupon it radiates to infinity.

\section{Formulas for amplitudes\label{sec-amplitudes}}

Given that wave propagation in layered media has been studied for more than half a century, it is curious that exact formulas for amplitudes have been obtained only recently.   Perhaps exact formulas for general amplitudes were viewed as being too cumbersome and unwieldy to be of practical use.  We show in the present section that on the contrary, amplitude formulas have a rich structure.  As a function of reflection coefficients, each amplitude is a tensor product of univariate functions that we call amplitude factors.   These have a simple expression in terms of classical Jacobi polynomials $P^{(\alpha,\beta)}_n$: reflection amplitude factors correspond to $\beta=1$, while transmission factors correspond to $\beta=0$.    The formulas obtained in $\cite{Gi:Comb2013}$ are not in factored form, but they provide an essential preliminary result that serves as a starting point for the analysis presented here.   

\subsection{Amplitude polynomials}

To begin, we fix notation with a pair of definitions. 
\begin{defn}[Amplitude factors]\label{defn-factor}
Let $(p,q)\in\integer^2$.
If $
\mbox{min}\{p,q\}<0,
$
set $f^{(p,q)}(x)=0$ and $g^{(p,q)}(x)=0$.   Set $f^{(0,0)}(x)=1$.  If $p>0$, set $f^{(p,0)}(x)=x^p$ and $f^{(0,p)}(x)=0$.  If 
$
\mbox{min}\{p,q\}\geq1,
$
set 
\[
f^{(p,q)}(x)=\sum_{j=1}^{{\rm min}\{p,q\}}(-1)^{q-j}\binom{p}{j}\binom{q-1}{j-1}x^{p+q-2j}(1-x^2)^j.
\]
If $\mbox{min}\{p,q\}\geq 0$, set
\[
g^{(p,q)}(x)=\sqrt{1-x^2}\sum_{j=0}^{{\rm min}\{p,q\}}(-1)^{q-j}\binom{p}{j}\binom{q}{j}x^{p+q-2j}(1-x^2)^j.
\]
We call the functions $f^{(p,q)}$ and $g^{(p,q)}$ amplitude factors.  
\end{defn}

\begin{defn}[Amplitude polynomials]\label{defn-amplitude}
For each $n\geq 1$, each lattice point $k=(k_0,\ldots,k_n)\in\integer^{n+1}$ and variables $x=(x_0,\ldots,x_n)$, 
set
\[
\begin{split}
a(x,k)&=\delta_{1 k_0}\,x_n^{k_n}\prod_{j=0}^{n-1}f^{(k_j,k_{j+1})}(x_j),\\
b(x,k)&=\delta_{0 k_0}\,\sqrt{1-x_n^2}\,x_n^{k_n}\prod_{j=0}^{n-1}g^{(k_j,k_{j+1})}(x_j).
\end{split}
\]
The $a(x,k)$ and $b(x,k)$, viewed as functions of $x_0,\ldots,x_n$ indexed by $k$, will be called amplitude polynomials and amplitude quasi-polynomials, respectively.  
\end{defn}

(The Kronecker delta
\[
\delta_{j k_0}=\left\{
\begin{array}{cc}
1&\mbox{ if }k_0=j\\
0&\mbox{ otherwise }
\end{array}
\right. 
\]
is included to allow arbitrary $k\in\integer^{n+1}$, which helps to simplify later formulas.)

The following result shows that amplitude factors $f^{(p,q)}$ play a role in the time domain that is roughly analogous to that of disk automorphisms in the frequency domain representation (\ref{G-hat}).  Given dimension $n\geq 1$, we use the notation $\mathbb{1}=(1,1,\ldots,1)\in\integer^{n+1}$.   
\begin{thm}\label{thm-main}
For each pair $(\tau^\prime,R)$, where 
\[
\tau^\prime=(\tau_0,\ldots,\tau_{n+1})\in\real_{>0}^{n+2},\quad\quad
R=(R_0,\ldots,R_n)\in[-1,1]^{n+1}
\]
and $\tau=(\tau_0,\ldots,\tau_{n})$, the corresponding reflection Green's function is 
\begin{equation}
G^{(\tau,R)}(t)=\sum_{\quad k\in\integer^{n+1}}a(R,k)\delta(t-\langle k,\tau\rangle).\label{reflection}
\end{equation}
The transmission Green's function is 
\begin{equation}
H^{(\tau^\prime,R)}(t)=\sum_{\quad k\in\integer^{n+1}}b(R,k)\delta\bigl(t-{\scriptstyle\frac{1}{2}}\tau_{n+1}-\langle k+{\scriptstyle\frac{1}{2}}\mathbb{1},\tau\rangle\bigr).\label{transmission}
\end{equation}
\end{thm}

\begin{pf} 
Let $\mathfrak{L}_n$ denote the set of all $(k_0,k_1,\ldots,k_n)\in\integer_+^{n+1}$ such that 
\[
k_0=1\mbox{ and }\forall j\leq n-1,\,k_j=0\Rightarrow k_{j+1}=0.\]
It follows from Definitions~\ref{defn-amplitude} and \ref{defn-factor} that $a(x,k)$ is not identically zero only if $k\in\mathfrak{L}_n$.   The following result is proved in \cite[Theorem~2.1]{Gi:Comb2013}.  Given $k=(k_0,\ldots,k_n)\in\integer^{n+1}$, let $\tilde{k}$ denote the left shift
$\tilde{k}=(k_1,\ldots,k_n,0)$; and given $x\in\real^{n+1}$ let $x^k$ denote the standard multi-index notation for $\prod_{j=0}^nx_j^{k_j}$.  Then 
\[
G^{(\tau,R)}(t)=\sum_{k\in\mathfrak{L}_n}\alpha_k\,\delta(t-\langle k,\tau\rangle),
\]
where, setting $u=\min\{\mathbb{1},\tilde{k}\}$, and letting $V(k)$ denote the set of $b\in\integer^{n+1}$ such that $u\leq b\leq\min\{k,\tilde{k}\}$,
\begin{equation}\label{aRk}
\alpha_k=\sum_{b\in V(k)}\binom{k}{b}\binom{\tilde{k}-u}{b-u}(-R)^{\tilde{k}-b}R^{k-b}T^{2b}.
\end{equation}
Here $T_j=\sqrt{1-R_j^2}$, and $T=(T_0,T_1,\ldots,T_n)$.   Also, 
$\binom{k}{b}$ and $\binom{\tilde{k}-u}{b-u}$ are multinomial coefficients, so that, for example, 
\[
\binom{k}{b}=\prod_{j=1}^n\binom{k_j}{b_j}.
\]
Let $k\in\mathfrak{L}_n$ and let $m$ denote the largest index such that $k_m\neq0$.  By Definitions~\ref{defn-amplitude} and \ref{defn-factor}, 
\[
\begin{split}
&a(R,k)=\\
&\delta_{1 k_0}\,R_m^{k_m}\prod_{s=0}^{m-1}\sum_{j=1}^{{\rm min}\{k_s,k_{s+1}\}}(-1)^{k_{s+1}-j}\binom{k_s}{j}\binom{k_{s+1}-1}{j-1}R_s^{k_s+k_{s+1}-2j}(1-R_s^2)^j.
\end{split}
\]
Expansion of this product yields precisely (\ref{aRk}).  Thus $\alpha_k=a(R,k)$, proving (\ref{reflection}).  

According to \cite[Theorem~3.1]{Gi:Comb2013}, the transmission Green's function has the form
\[
H^{(\tau^\prime,R)}(t)=\sum_{k\in\{0\}\times\integer^n_+}\beta_k\,\delta\bigl(t-{\scriptstyle\frac{1}{2}}|\tau^\prime|-{\scriptstyle\frac{1}{2}}\langle k,\tau\rangle\bigr)
\]
where for each $k\in\{0\}\times\integer^n_+$, the amplitude $\beta_k$ is given by the formula\begin{equation}\label{bRk}
\beta_k=\sum_{0\leq m\leq\min\{k,\tilde{k}\}}\binom{k}{m}\binom{\tilde{k}}{m}(-R)^{\tilde{k}-m}R^{k-m}T^{2m+\mathbb{1}}.
\end{equation}
As in the case of reflection, the formula for $\beta_k$ is easily reconciled with the formula for $b(R,k)$ given in Definitions~\ref{defn-amplitude} and \ref{defn-factor}, proving (\ref{transmission}).   
\end{pf}

The energy relation (\ref{conservation}) yields the following useful corollary to Theorem~\ref{thm-main}.
\begin{cor}\label{cor-conservation}
Let $x\in[-1,1]^{n+1}$ for some $n\geq 1$.  Then 
\[
\sum_{k\in\integer^{n+1}}a(x,k)^2+b(x,k)^2=1.
\]
\end{cor}
\begin{pf} Choose $\tau^\prime=(\tau_0,\ldots,\tau_{n+1})\in\real_{>0}^{n+2}$ such that the numbers $\tau_j$ are linearly independent over the integers.   Then the arrival times $\langle k,\tau\rangle$ occuring in (\ref{reflection}) are all distinct, as are the arrival times in (\ref{transmission}).   The conservation relation (\ref{conservation}) therefore implies the statement of the corollary. 
\end{pf}

\subsection{Covering amplitude\label{sec-covering}}

Let $\lfloor\cdot\rfloor$ denote the floor (i.e., least integer) function, interpreted entrywise on vectors, so that for 
\[
x=(x_0\,x_1,\ldots,x_{n})\in\real^{n+1},\quad\lfloor x\rfloor=\bigl(\lfloor x_0\rfloor,\ldots,\lfloor x_{n}\rfloor\bigr).
\]
Provided one restricts reflection coefficients to the range $[-1,1)$, i.e. disallowing the value 1, one can regard the amplitude polynomials in $n+1$ variables as patches of a larger map defined as follows.
\begin{defn}[Covering amplitude]\label{defn-covering}
Write $\tilde{x}=x-2\lfloor (x+1)/2\rfloor$.  For each $x\in\real^{n+1}$, set 
$
\mathfrak{a}_n(x)=a\bigl(\tilde{x},\lfloor (x+1)/2\rfloor\bigr), 
$
so that 
\[
\mathfrak{a}_n(x+2k)=a(x,k)\quad\bigl(k\in\integer^{n+1},\;x\in[-1,1)^{n+1}\bigr).
\]
We call $\mathfrak{a}_n:\real^{n+1}\rightarrow [-1,1]$ the covering amplitude.   
\end{defn}
In terms of the covering amplitude, the set of amplitudes $\{a_j\}$ in (\ref{train}) is precisely $\mathfrak{a}_n(R+2\integer^{n+1})$.  Thus the reflectivity $R$ can be viewed as a translation---the covering amplitude $\mathfrak{a}_n$ is sampled on precisely this translate of the twice-integer lattice to yield the reflection amplitudes.

\subsection{Orthogonal polynomials}

To bring to light a connection that amplitude factors have to the Poincar\'e disk (see Section~\ref{sec-backward}), we mimic a construction used by Szeg\H{o}, but with the measure induced by the Poincar\'e metric instead of the usual Lebesgue measure.   In \cite[\S 5,6]{Sz:1918}, Szeg\H{o} studied the classes of polynomials obtained by orthogonalizing the sequence 
\begin{equation}\label{sequence}
|x|^{\alpha+1/2}x^n\quad(n\in\integer_+)
\end{equation}
with respect to Lebesgue measure on the interval $[-1,1]$, for fixed values of $\alpha\in\integer_+$.   Viewing $[-1,1]$ as a diameter of the Poincar\'e disk, the distance in the Poincar\'e metric between 0 and $\pm x$, for $0<x<1$, is 
\[
\frac{1}{2}\log\frac{1+|x|}{1-|x|},
\]
which induces the measure
\begin{equation}\label{Poincare}
\frac{dx}{1-x^2}
\end{equation}
on $[-1,1]$.  The sequence (\ref{sequence}) cannot  be orthogonalized with respect to (\ref{Poincare}), because the given monomials are not integrable.  The situation is remedied by replacing (\ref{sequence}) with the integrable sequence
\begin{equation}\label{integrable}
|x|^{\alpha+1/2}(1-x^2)x^{n}\quad(n\in\integer_+).
\end{equation}

 Orthogonalization of (\ref{integrable}) with respect to (\ref{Poincare}) produces a sequence of functions
 \[
 |x|^{\alpha+1/2}(1-x^2)q^{(\alpha)}_{n}\quad(n\in\integer_+),
 \]
where each $q^{(\alpha)}_{n}$ is a monic polynomial of degree $n$.  The polynomials 
\[
x^\alpha(1-x^2)q^{(\alpha)}_{2n}(x)
\]
are proportional to amplitude factors, and hence they play a dual role to disk automorphisms in the representation of $G$.  (Note that only the $q^{(\alpha)}_{2n}$ having even degree occur in amplitude factors.)  
\begin{lem}\label{lem-Jacobi}For $\alpha,\beta>-1$ and $n\in\integer_+$, let $P^{(\alpha,\beta)}_n$ denote the classical Jacobi polynomial, 
\[
P^{(\alpha,\beta)}_n(x)=\sum_{j=0}^n\binom{n+\alpha}{n-j}\binom{n+\beta}{j}\Bigl(\frac{x-1}{2}\Bigl)^j\Bigl(\frac{x+1}{2}\Bigl)^{n-j}.
\]
For each $\alpha\in\integer_+$, 
\[
q^{(\alpha)}_{2n}(x)\propto P^{(\alpha,1)}_n(1-2x^2).
\]
\end{lem}
\begin{pf}  By definition, the polynomials $q^{(\alpha)}_n(x)$ are orthogonal with respect to the measure 
\[
x^{2\alpha+1}(1-x^2)\,dx.
\]
The Jacobi polynomials $P^{(\alpha,1)}_n(z)$ are orthogonal with respect to  
\[
\left(\frac{1-z}{2}\right)^\alpha \left(\frac{1+z}{2}\right)\,dz,
\]
which, upon changing variables to $z=1-2x^2$, becomes
\[
-4x^{2\alpha+1}(1-x^2)\,dx.
\]
Therefore, matching degrees, $q^{(\alpha)}_{2n}(x)\propto P^{(\alpha,1)}_n(1-2x^2)$.   
\end{pf}

\begin{thm}\label{thm-factor-Jacobi}Let $(p,q)\in\integer^2$.  Set $\alpha=|p-q| $ and $m=\mbox{min}\{p,q\}-1$.   If $m\geq 0$, then
\begin{equation}\label{f-Jacobi}
f^{(p,q)}(x)=\left\{\begin{array}{cc}
(-x)^{\alpha }(1-x^2)P^{(\alpha ,1)}_m(1-2x^2)&\mbox{ if }p\leq q\\
\rule{0pt}{20pt}\displaystyle\frac{p}{q}\,x^{\alpha }(1-x^2)P^{(\alpha ,1)}_m(1-2x^2)&\mbox{ if }p> q\\
\end{array}\right.
\end{equation}
whereby $f^{(p,q)}(x)\propto x^\alpha(1-x^2)q^{(\alpha)}_{2n}(x)$.   Set $n=\mbox{min}\{p,q\}$.  If $n\geq 0$ then
\begin{equation}\label{g-Jacobi}
g^{(p,q)}(x)=\left\{\begin{array}{cc}
(-x)^{\alpha }\sqrt{1-x^2}\;P^{(\alpha ,0)}_n(1-2x^2)&\mbox{ if }p\leq q\\
\rule{0pt}{20pt}\displaystyle x^{\alpha }\sqrt{1-x^2}\;P^{(\alpha ,0)}_n(1-2x^2)&\mbox{ if }p> q\\
\end{array}\right..
\end{equation}
\end{thm}
\begin{pf}
The formulas (\ref{f-Jacobi}) and (\ref{g-Jacobi}) can be verified directly by comparing Definition~\ref{defn-factor} to the standard formula for Jacobi polynomials appearing in Lemma~\ref{lem-Jacobi}.  The fact that $f^{(p,q)}(x)\propto x^\alpha(1-x^2)q^{(\alpha)}_{2n}(x)$ is then a consequence of Lemma~\ref{lem-Jacobi}.   
\end{pf}

\section{Polydisk functions\label{sec-polydisk}}

We recall some basic facts about Hardy space on the polydisk, citing \cite{Ru:1969} as a general reference.  Let $z=(z_0,\ldots,z_n)$ denote a complex $n+1$-tuple.  The Hardy space $H^2(\ddd^{n+1})$ consists of all holomorphic functions on the polydisk
\[
\varphi:\ddd^{n+1}\rightarrow\complex
\]
whose Taylor expansions
\begin{equation}\label{Taylor}
\varphi(z)=\sum_{k\in\integer^{n+1}_+}a_k\,z^k,
\end{equation}
have coefficients in $\ell_2(\integer^{n+1}_+)$, meaning that 
\[
\sum_{k\in\integer^{n+1}_+}|a_k|^2<\infty.
\]
See \cite[p.~50]{Ru:1969}.  By Fatou's Theorem, such functions extend almost everywhere to the distinguished boundary of the polydisk, the torus $\ttt^{n+1}$.   We use the same symbol $\varphi$ for the boundary function as for the original function.   The boundary function $\varphi:\ttt^{n+1}\rightarrow\complex$ belongs to $L^2(\ttt^{n+1})$ and has $L^2$ norm 
\begin{equation}\label{norm}
||\varphi||_2=\left(\sum_{k\in\integer^{n+1}_+}|a_k|^2\right)^{1/2},
\end{equation}
with respect to which $H^2(\ddd^{n+1})$ is a Hilbert space.   (Indeed, $H^2(\ddd^{n+1})$ can be realized as the closed subspace of $L^2(\ttt^{n+1})$ consisting of those functions whose Fourier coefficients are zero for frequencies $e^{i\langle k,\xi\rangle}$ where $k\not\in\integer^{n+1}_+$.)
A function $\varphi\in H^2(\ddd^{n+1})$ is by definition inner if 
\[
|\varphi(e^{i\xi_0},\ldots,e^{i\xi_n})|=1
\]
almost everywhere on $\ttt^{n+1}$; see \cite[Chapter~5]{Ru:1969}.   Inner functions play a central role in the following sections.

\subsection{Polydisk families \label{sec-families}}
By Corollary~\ref{cor-conservation} and the fact that $a(x,k)$ and $b(x,k)$ are identically zero if $k\not\in\integer^{n+1}_+$, it is natural to view the amplitude polynomials as coefficients of functions in $H^2(\ddd^{n+1})$.   

\begin{defn}\label{defn-polydisk} Let $n\geq 1$ be an integer, and let $z=(z_0,\ldots,z_n)\in\ddd^{n+1}$.  For each $x=(x_0,\ldots,x_n)\in[-1,1]^{n+1}$, set 
\[
\begin{split}
\varphi_x(z)=&\sum_{k\in\integer^{n+1}}a(x,k)z^k,\\
\psi_x(z)=&\sum_{k\in\integer^{n+1}}b(x,k)z^k.
\end{split}
\]
We call $\varphi_x$ and $\psi_x$ polydisk functions, and we call $\{\varphi_x\}$ and $\{\psi_x\}$ the reflection and transmission polydisk families, respectively.   
\end{defn}

 Note that the polydisk families are uniformly bounded and hence normal, by Montel's theorem \cite[Theorem~1.5]{Oh:2002}.

\subsection{Almost periodic functions}

A key fact is that the restriction of a polydisk function to a generic line on the torus is almost periodic in the sense of Besicovitch and its Besicovitch norm coincides with the $H^2$ norm of the original function.  

In detail, a function $f:\real\rightarrow\complex$ is almost periodic in the sense of Besicovitch if and only if it has a Fourier series representation of the form 
\[
f(\omega)\sim\sum_{j=1}^\infty\alpha_je^{i\lambda_j\omega}\quad\mbox{ with }\quad\sum_{j=1}^\infty|\alpha_j|^2<\infty,
\]
and where the $\lambda_j$ are real and distinct.  For such a function the Besicovitch norm $||f||_B$ has two representations, 
\begin{equation}\label{B-norm}
\begin{split}
||f||_B^2&=\lim_{T\rightarrow\infty}\frac{1}{2T}\int_{-T}^T|f(\omega)|^2\,d\omega\\
&=\sum_{j=1}^\infty|\alpha_j|^2.
\end{split}
\end{equation}
See \cite[Chapter~II]{Be:1955}.  Since the Taylor coefficients of an arbitrary $\varphi\in H^2(\ddd^{n+1})$ are square summable, the restriction 
\[
\varphi(e^{i\omega\tau_0},\ldots,e^{i\omega\tau_n})=\sum_{k\in\integer^{n+1}_+}a_k\,e^{i\langle k,\tau\rangle\omega}
\]
of $\varphi(z)=\sum_{k\in\integer^{n+1}_+}a_k\,z^k$ to the line on the torus
\[
\ell_\tau=\left\{ (e^{i\tau_0\omega},\ldots,e^{i\tau_n\omega})\,\left|\,\omega\in\real\right.\right\}
\]
is almost periodic, provided the real numbers $\langle k,\tau\rangle$ are distinct.  This is true in particular if the components of $\tau=(\tau_0,\ldots,\tau_n)$ are linearly independent over the integers---which is the generic case.   Moreover, by (\ref{B-norm}), the Besicovitch norm of the generic restriction agrees with the Hardy space norm of the original function:
\begin{equation}\label{B-H}
\lVert\,\varphi|_{\ell_\tau}\rVert_B=\lVert\varphi\rVert_2=\left(\sum_{k\in\integer^{n+1}_+}|a_k|^2\right)^{1/2}.
\end{equation}

Applying the Fourier transform to the representation of $G^{(\tau,R)}$ from Theorem~\ref{thm-main} yields the formula
\begin{equation}\label{fourier}
\widehat{G^{(\tau,R)}}(\omega)=\sum_{\quad k\in\integer^{n+1}}a(R,k)\,e^{i\omega\langle k,\tau\rangle},
\end{equation}
the restriction of $\varphi_R$  to the line $\ell_\tau$.   Therefore the Besicovitch norm of $\widehat{G^{(\tau,R)}}$ is the $H^2$ norm of $\varphi_R$, provided $\tau$ is generic.

Among other things, the representation (\ref{fourier}) makes clear the dependence of $\widehat{G^{(\tau,R)}}$ on $R$ and $\tau$:  the reflection coefficients determine $\varphi_R$; and $\tau$ determines a line on the torus.  This is in contrast to the backward recurrence (\ref{G-hat}), in which it is not a priori clear how to disentangle the role of $\tau$ from that of $R$.   

Lines on the torus correspond in the time domain to lattice projections, which is the subject of Section~\ref{sec-lattice}.   The correspondence between functions on the torus, almost periodic functions, and lattice projections also arises in connection with quasi crystals, as illustrated in the work of Moody et al.~\cite{MoNePa:2008}.

\subsection{An energy estimate}

We return now to the polydisk families introduced in Section~\ref{sec-families}, using the formulas for amplitude polynomials to establish a basic estimate on the $H^2$ norm of a polydisk function $\varphi_x$.     

\begin{thm}\label{thm-energy}   For every $x\in[-1,1]^{n+1}$, 
$
\max\limits_{0\leq j\leq n}\{|x_j|\}\leq ||\varphi_x||_{2}\leq 1.
$
\end{thm}
\begin{pf}
The inequality $||\varphi_x||_{2}\leq 1$ follows from Definition~\ref{defn-polydisk}, the formula (\ref{norm}) and Corollary~\ref{cor-conservation}.

It follows directly from Definitions~\ref{defn-amplitude} and \ref{defn-factor} that if $k\in\integer^{n+1}$ has the property that $k_{j+1}=0$ for some $0\leq j\leq n-1$, then 
\begin{equation}\label{compatibility}
a\bigl((x_0,\ldots,x_j),(k_0,\ldots,k_j)\bigr)=a\bigl((x_0,\ldots,x_n),(k_0,\ldots,k_n)\bigr).
\end{equation}
The representation (\ref{g-Jacobi}) of Theorem~\ref{thm-factor-Jacobi} shows that 
\[
|g^{(p,q)}|=|g^{(q,p)}|,
\]
from which it follows that 
\begin{equation}\label{b-symmetry}
b\bigl((x_0,\ldots,x_j),(0,k_1,\ldots,k_j)\bigr)^2=b\bigl((x_j,\ldots,x_0),(0,k_j\ldots,k_1)\bigr)^2.
\end{equation}
Note also that for $k\in\integer^{n+1}$, $b(x,k)$ is identically zero unless $k\in\{0\}\times\integer^{n}_+$.   Since $\integer^n_+$ is invariant under permutations, it follows in particular that 
\[
||\psi_{(x_0,\ldots,x_j)}||_2^2=\sum_{k\in\integer^{j+1}}b\bigl((x_0,\ldots,x_j),k\bigr)^2
\]
is invariant under the permutation on $\integer^{j+1}$,
\begin{equation}\label{perm}
(k_0,k_1,k_2,\ldots,k_j)\mapsto(k_0,k_j,k_{j-1},\ldots,k_1).
\end{equation}
Invariance of (\ref{perm}) combined with (\ref{b-symmetry}) yields that 
\begin{equation}\label{psi-symmetry}
||\psi_{(x_0,\ldots,x_j)}||_2=||\psi_{(x_j,\ldots,x_0)}||_2,
\end{equation}
which, by Corollary~\ref{cor-conservation}, implies that 
\begin{equation}\label{phi-symmetry}
||\varphi_{(x_0,\ldots,x_j)}||_2=||\varphi_{(x_j,\ldots,x_0)}||_2.
\end{equation}
By Definitions~\ref{defn-amplitude} and \ref{defn-factor}, $a(x,k)=x_0$ 
for $k=(1,0,\ldots,0)$; this implies that $||\varphi_{(x_0,\ldots,x_j)}||_2\geq |x_0|$, which, by (\ref{phi-symmetry}), implies that 
\begin{equation}\label{j-estimate}
||\varphi_{(x_0,\ldots,x_j)}||_2\geq |x_j|.
\end{equation}
The comparison (\ref{compatibility}) between dimensions shows that if $j\leq n$ then
\[
||\varphi_{(x_0,\ldots,x_n)}||_2\geq||\varphi_{(x_0,\ldots,x_j)}||_2.
\]
Combined with (\ref{j-estimate}) this proves $||\varphi_{(x_0,\ldots,x_n)}||_2\geq|x_j|$.   Since this is true for every $j\leq n$, the theorem follows.\end{pf}

\subsection{Inner polydisk functions\label{sec-inner}} 

\begin{cor}\label{cor-inner}
If $x\in[-1,1]^{n+1}$ is a boundary point, then $\varphi_x$ is inner.   
\end{cor}
\begin{pf}
A given $x\in[-1,1]^{n+1}$ is a boundary point when 
\[
\max\limits_{0\leq j\leq n}\{|x_j|\}=1,
\]
 in which case Theorem~\ref{thm-energy} forces $||\varphi_x||_{2}=1$.   The restriction of $\varphi_x$ to any line on the torus $\ell_\tau$ is the Fourier transform (\ref{fourier}) of a reflection Green's function and so satisfies
\begin{equation}\label{torus-norm}
|\varphi_x(z)|\leq 1
\end{equation}
for every $z\in\ell_\tau$.   Since $\tau$ can be chosen to pass through any point on the torus, it follows that (\ref{torus-norm}) holds for every $z\in\ttt^{n+1}$.  Using the representation of $\varphi_x$ in $L^2(\ttt^{n+1})$, the inequality (\ref{torus-norm}) and the fact that $||\varphi_x||_{2}=1$  imply that $|\varphi_x(z)|=1$ almost everywhere on $\ttt^{n+1}$, proving that $\varphi_x$ is inner.   \end{pf}

\begin{lem}\label{lem-inner}
Let $x\in[-1,1]^{n+1}$ for some $n\geq 1$, and set $\widetilde{\varphi}=\varphi_{(x_0,\ldots,x_{n-1},1)}$.   Then $\widetilde{\varphi}$ is inner, and 
\[
||\varphi_x-\widetilde{\varphi}||_2\leq \min\limits_{0\leq j\leq n-1}2\sqrt{1-x_j^2}.
\]
\end{lem}
\begin{pf}
The polydisk function $\widetilde{\varphi}$ is inner by Corollary~\ref{cor-inner}.   Let 
\[
\pi_j:H^2(\ddd^{n+1})\rightarrow H^2(\ddd^{n+1})
\]
 denote the orthogonal projection
\[
\sum_{k\in\integer^{n+1}_+}c_k\,z^k\quad\mapsto\sum_{\stackrel{k\in\integer^{n+1}_+}{k_{j+1}=\cdots=k_{n+1}=0}}c_k\,z^k.
\]
Observe that for $0\leq j\leq n-1$,
$
\pi_j\widetilde{\varphi}=\pi_j\varphi_x,
$
by (\ref{compatibility}), and that 
\[
||\pi_j\varphi_x||_2^2\geq x_j^2,
\]
by Theorem~\ref{thm-energy}.   
By orthogonality of $\pi_j$, 
\[
||\varphi_x-\pi_j\varphi_x||_2^2=||\varphi_x||_2^2-||\pi_j\varphi_x||_2^2\leq 1-x_j^2
\]
and
\[
||\widetilde{\varphi}-\pi_j\varphi_x||_2^2=||\widetilde{\varphi}||_2^2-||\pi_j\varphi_x||_2^2\leq 1-x_j^2,
\]
from which it follows by the triangle inequality that 
\[
||\widetilde{\varphi}-\varphi_x||_2\leq2\sqrt{1-x_j^2}.
\]
Since this is true for each $0\leq j\leq n-1$, the lemma follows.  
\end{pf}

\begin{thm}\label{thm-probability}
Let $0<\epsilon<2$, and let $x\in[-1,1]^{n+1}$ be chosen randomly with respect to the uniform distribution on $[-1,1]^{n+1}$.  The probability that 
\[
||\varphi_x-\widetilde{\varphi}||_2\leq\epsilon
\]
for some inner function $\widetilde{\varphi}$, is at least 
\[1-\Bigl(1-(\epsilon/2)^2\Bigr)^{n/2}.
\]
\end{thm}
\begin{pf}
Let $r=\max\limits_{0\leq j\leq n-1}|x_j|$, so that 
\[
\min\limits_{0\leq j\leq n-1}2\sqrt{1-x_j^2}\leq\epsilon\quad\mbox{ if and only if }\quad2\sqrt{1-r^2}\leq\epsilon.
\]
The proportion of $[-1,1]^{n+1}$ for which the latter inequality is satisfied is precisely \[
1-\sqrt{1-(\epsilon/2)^2}^{\,n}.
\]  
By Lemma~\ref{lem-inner}, the condition 
\[
\min\limits_{0\leq j\leq n-1}2\sqrt{1-x_j^2}\leq\epsilon
\]
guarantees that $\varphi_x$ is within distance $\epsilon$ of an inner function.   
\end{pf}

By the estimate in Theorem~\ref{thm-probability}, the probability of $\varphi_x$ being within distance $\epsilon$ of an inner function approaches 1 as $n\rightarrow\infty$.    In this sense the polydisk functions are approximately inner.

\section{The power spectrum of the Green's function\label{sec-power}}

Roughly speaking, the fact that the reflection polydisk family $\{\varphi_x\}$ consists of functions that are approximately inner (with high probability as $n\rightarrow\infty$) implies that for a large number of layers,
\[
\left|\widehat{G^{(\tau,R)}}(w)\right|\cong 1.
\]
The purpose of the present section is to formulate this result quantitatively.      

\begin{thm}\label{thm-unity}
Let $0<\epsilon<2$.  If the entries of $\tau\in\real^{n+1}_{>0}$ are linearly independent over the integers,  then the probability that a randomly chosen $R\in[-1,1]^{n+1}$ satisfies
\[
\lim_{T\rightarrow\infty}\frac{1}{2T}\int_{-T}^T\left(1-\left|\widehat{G^{(\tau,R)}}(\omega)\right|\right)^2\,d\omega\leq\epsilon^2
\]
is at least 
\[1-\Bigl(1-(\epsilon/2)^2\Bigr)^{n/2}.
\]
\end{thm}
\begin{pf}
Let $\tau$ have linearly independent entries over $\integer$, write 
\[
\widetilde{R}=(R_0,\ldots,R_{n-1},1),
\]
 and set $\widetilde{\varphi}=\varphi_{\widetilde{R}}$.   Observe that $\widehat{G^{(\tau,\widetilde{R})}}=e^{i\omega\tau_n}$ has constant modulus 1 by the backward recurrence formulas (\ref{automorphism}) and (\ref{G-hat}).  Thus 
\[
\begin{split}
\lim_{T\rightarrow\infty}\frac{1}{2T}\int_{-T}^T&\left(1-\left|\widehat{G^{(\tau,R)}}(\omega)\right|\right)^2\,d\omega\\
&=
\lim_{T\rightarrow\infty}\frac{1}{2T}\int_{-T}^T\left(\left|\widehat{G^{(\tau,\widetilde{R})}}(\omega)\right|-\left|\widehat{G^{(\tau,R)}}(\omega)\right|\right)^2\,d\omega\\
&\leq
\lim_{T\rightarrow\infty}\frac{1}{2T}\int_{-T}^T\left|\widehat{G^{(\tau,\widetilde{R})}}(\omega)-\widehat{G^{(\tau,R)}}(\omega)\right|^2\,d\omega\\
&=\lVert \widehat{G^{(\tau,\widetilde{R})}}-\widehat{G^{(\tau,R)}}\rVert^2_B\\
&=\lVert \widetilde{\varphi}-\varphi_R\rVert^2_2\,\mbox{ by (\ref{B-H})}.
\end{split}
\]
The result then follows from Lemma~\ref{lem-inner} and the proof of Theorem~\ref{thm-probability}.  \end{pf}

Theorem~\ref{thm-unity} predicts that for a sufficiently large number of layers, the Fourier transform of the reflection Green's function has approximately unit modulus, but the estimate it uses is rather weak.   In numerical experiments this phenomenon manifests itself already for media having on the order of 10-12 layers, much sooner than necessitated by the theorem.

\section{Arrival time inversion\label{sec-arrival}}

The basic problem under consideration in the present section is to recover $\tau$ from the arrival times $\langle k,\tau\rangle$ comprising the support of $G^{(\tau,R)}(t)$ (as in Theorem~\ref{thm-main}), up to a cutoff time $T=\tau_0+\cdots+\tau_n$.  (Note that this is the minimum $T$ for which it's possible to determine $\tau_n$.)   Once the travel time vector $\tau$ is known, then the amplitude polynomials can be brought to bear on the problem of recovering $R$ from amplitude data, which is treated in Section~\ref{sec-localized}.

\subsection{The lattice projection problem\label{sec-lattice}}  

By the representation (\ref{reflection}), recovering $\tau$ from arrival times is equivalent to recovering a linear functional from its values on part of the integer lattice.  The following notation will be used.  Let $L_\tau$ denote the linear functional corresponding to a given $\tau\in\real^{n+1}_{>0}$, so that for $x\in\real^{n+1}$,
\[
L_\tau(x)=\langle x,\tau\rangle.
\]
The symbol $\lll_n$ denotes the subset of $\integer^{n+1}_+$ supported on initial intervals; i.e. for every $k\in\integer^{n+1}_+$, $k\in\lll_n$ if and only if
\[
k_j=0\Rightarrow k_{j+1}=0\quad  (0\leq j\leq n-1).
\]
Let $\one\in\real^{n+1}$ denote the constant vector, each of whose entries is $1$.  The same symbol $\one$ will be used for all dimensions; it will be clear from context which value of $n$ is appropriate.   Given $\tau\in\real^{n+1}_{>0}$, let $\lll^\tau$ denote the set 
\[
\lll^\tau=\bigl\{k\in\lll_n\,\bigl|\,0<\langle k,\tau\rangle\leq\langle\one,\tau\rangle\bigr\}.
\]
Finally, it will be convenient to represent the set $L_\tau(\lll^\tau)$ by its elements ordered in a vector, denoted $\Phi(\tau)$.   That is, $\sigma=\Phi(\tau)$ means that 
\[
\sigma_1<\cdots<\sigma_d
\mbox{ and }
\bigl\{\sigma_1,\ldots,\sigma_d\bigr\}=L_\tau(\lll^\tau).
\] 
This defines a mapping 
\[
\Phi:\bigcup_{n=1}^\infty\real^{n+1}_{>0}\rightarrow \bigcup_{n=1}^\infty\real^{n+1}_{>0}.
\]
Note that the mapping $\Phi$ commutes with multiplication by a positive scalar: for any $\alpha>0$ and $\tau\in\bigcup_{n=1}^\infty\real^{n+1}_{>0}$, 
\begin{equation}\label{commutes}
\alpha\Phi(\tau)=\Phi(\alpha\tau).
\end{equation}
The precise problem of interest is to recover $\tau$ from $\Phi(\tau)$.  Letting $\ell_\tau$ denote the line in $\real^{n+1}$ spanned by $\tau$, the problem of recovering $\tau$ from $\Phi(\tau)$ is called a \emph{lattice projection problem} in reference to the fact that the entries of $\Phi(\tau)$ are the norms of the orthogonal projections of the lattice points $\lll^\tau$ onto $\ell_\tau$, rescaled by $||\tau||$.

\subsection{Factorization of $\Phi(\tau)$ when $L_\tau$ is injective on $\lll_\tau$}
For $x\in\real^{n+1}$, let $x^\perp$ denote its orthogonal complement, the hyperplane
\[
x^\perp=\bigl\{y\in\real^{n+1}\,\bigl|\,\langle y,x\rangle=0\bigr\}.
\]
Let $H_x^+$ denote the closed half space bounded by $x^\perp$ on which $L_x$ is non-negative,
\[
H_x^+=\bigl\{y\in\real^{n+1}\,\bigl|\,\langle y,x\rangle\geq0\bigr\}.
\]
Given $\tau\in\real^{n+1}_{>0}$ and $k\in\lll_n$, observe that 
$
k\in\lll^\tau$ if and only if $\tau\in H_{\mathbb{1}-k}^+.
$
The observation below follows directly from the established notation.   
\begin{prop}\label{prop-non-injective}
Given $\tau\in\real^{n+1}_{>0}$, the linear functional $L_\tau$ fails to be injective on $\lll^\tau$ if and only if 
$\tau$ belongs to 
\begin{equation}\label{half-hyperplane}
H_{\mathbb{1}-k}^+\cap(k^\prime-k)^\perp
\end{equation}
for some pair of distinct lattice points $k,k^\prime\in\lll_n$.
\end{prop}
Let $\hhh_n$ denote the union of all sets of the form $H_k^+\cap(k-k^\prime)^\perp$, where $k\neq k^\prime$ belong to $\lll_n$. Each of these sets is either empty, a half hyperplane, or a hyperplane (in the case $k^\prime=\mathbb{1}$).   Since $\hhh_n$ has measure zero in $\real^{n+1}$, it follows that for a generic set of $\tau\in\real^{n+1}_{>0}$, namely
\[
\ggg_n=\real^{n+1}_{>0}\setminus\hhh_n,
\]
the linear functional $L_\tau$ is injective on $\lll^\tau$.  Restricted to 
\[
\ggg=\bigcup_{n=1}^\infty\ggg_n,
\]
the lattice projection problem can be expressed as a factorization problem, as follows. 
For every $\tau\in\ggg_n$ (viewed as a column vector) there is a unique matrix $A$ whose rows belong to $\lll_n$ such that 
\begin{equation}\label{factorization}
A\tau=\Phi(\tau).
\end{equation}
This is a direct consequence of injectivity of $L_\tau$ on $\lll^\tau$; the rows of $A$ are simply the elements $k$ of $\lll^\tau$, ordered from top to bottom according to increasing value of $L_\tau(k)$.  Uniqueness of the factorization (\ref{factorization}) fails precisely when $L_\tau$ is not injective on $\lll^\tau$, i.e., when $\tau\in\cup_{n=1}^\infty\hhh_n$.\\[5pt]
\parbox{.95\textwidth}{
\textbf{Factorization problem}.   Let $\sigma\in\Phi(\ggg)$ be given.  Determine $N\geq1$, $\tau\in\real^{N+1}_{>0}$, and an integer matrix $A$ such that $\sigma=A\tau$, where the rows of $A$ belong to $\lll_N$.   
}

\subsection{The solution algorithm}
The following algorithm gives a direct method to compute the integer matrix $A$ and the vector $\tau$ from their product $\widetilde{\sigma}\in\Phi(\ggg)$, as required in the factorization problem.   In fact any \emph{primary subvector} $\sigma$ of $\widetilde{\sigma}\in\Phi(\ggg)$ suffices, as follows. 

\begin{defn}[Primary subvector]\label{defn-primary-subvector}   Let $\widetilde{\sigma}\in\Phi(\ggg_n)$ have factorization $A\tau$.   Any subvector $\sigma$ of $\widetilde{\sigma}$ (obtained by deleting entries) that includes all of the numbers $\tau_0+\cdots+\tau_j$, for $0\leq j\leq n$, is called a primary subvector of $\widetilde{\sigma}$.   
\end{defn}

\begin{alg}\label{alg-solution}
\underline{Input:}\\ Let $\sigma=\begin{pmatrix}\sigma_1\\ \vdots \\ \sigma_d\end{pmatrix}$ be a primary subvector of some $\widetilde{\sigma}\in\Phi(\ggg)$.\\[5pt]
\underline{Initial Step:}  Set $\tau_0=\sigma_1$, set $\tau^0=(\tau_0)$ (viewed as a $1\times 1$ vector), and set
\[
\lll^\tau_0=\bigl\{k\in\lll_0\,\bigl|\,\tau_0\leq k\tau_0\leq \sigma_d\bigr\}.
\]
Construct $\sigma^0$ from $\sigma$ by deleting the entries of $\sigma$ belonging to $L_{\tau^0}(\lll^\tau_0)$.\\[5pt]     
\underline{Continuing Step:}  Let $\tau^n=\begin{pmatrix}\tau_0\\ \vdots \\ \tau_n\end{pmatrix}$, $\lll^\tau_n$ and $\sigma^n$ be given.   If $\sigma^n=\emptyset$ then set $N=n$, $\tau=\tau^n$, and go to the output step.  Otherwise, set $\tau_{n+1}=\sigma^n_1-\langle\mathbb{1},\tau^n\rangle$, set $\tau^{n+1}=\begin{pmatrix}\tau_0\\ \vdots \\ \tau_{n+1}\end{pmatrix}$, and set 
\[
\lll^\tau_{n+1}=\bigl\{k\in\lll_{n+1}\,\bigl|\,k_{n+1}\geq 1\mbox{ and }
\langle k,\tau^{n+1}\rangle\leq\sigma_d\bigr\}.
\]
Construct $\sigma^{n+1}$ from $\sigma^n$ by deleting the elements of $L_{\tau^{n+1}}(\lll^\tau_{n+1})$ from $\sigma^n$.\\[5pt]
\underline{Output Step}: For $0\leq n\leq N$, extend each element of $\lll^\tau_n$ by a string of $N-n$ zeros to form $\widetilde{\lll}^\tau_n\subset\integer^{N+1}_+$.   Construct $A$ to be the $d\times (N+1)$ array whose rows consist of the elements $k$ of $\bigcup_{n=0}^N\widetilde{\lll}^\tau_n$, ordered such that $L_\tau(k)$ increases from the top row to the bottom.   Output the pair $(A,\tau)$.  \end{alg}

The fact that the pair $(A,\tau)$ solves the factorization problem for the given input $\sigma$ rests on the following straighforward proposition.
\begin{prop}\label{prop-tau}
The value $\tau_0+\cdots+\tau_{n+1}$ is the least element of 
\[
L_\tau(\lll^\tau)\setminus\bigcup_{j=0}^nL_{\tau^j}(\lll^\tau_j),
\]
for each $n$ in the range $0\leq n\leq N-1$.  
\end{prop}

\begin{thm}\label{thm-alg}
If $\sigma$ is a primary subvector of $\widetilde{\sigma}\in\Phi(\ggg)$ then the corresponding output $(A,\tau)$ of Algorithm~\ref{alg-solution} satisfies $\widetilde{\sigma}=A\tau$.   
\end{thm}
\begin{pf} Let $(\widetilde{A},\widetilde{\tau})$ be the unique pair such that $\widetilde{\sigma}=\widetilde{A}\widetilde{\tau}$, with $\widetilde{\tau}\in\ggg$.  If 
\[
(\tau_0,\ldots,\tau_n)=(\widetilde{\tau}_0,\dots,\widetilde{\tau_n})
\]
then Proposition~\ref{prop-tau} implies that in the Continuing Step, $\sigma_1^n$, which is the least element of $\sigma^n$, has the form 
\[
\sigma_1^n=\widetilde{\tau}_0+\cdots+\widetilde{\tau}_{n+1},
\]
so that $\sigma^n_1-\langle\mathbb{1},\tau^n\rangle$ is exactly $\widetilde{\tau}_{n+1}$.   It follows by induction that $\tau=\widetilde{\tau}$.   

Observe that $A\widetilde{\tau}=\widetilde{\sigma}$ by construction.   Therefore $\widetilde{A}=A$ since the equation $X\widetilde{\tau}=\widetilde{\sigma}$ has a unique solution by definition of $\ggg$.   
\end{pf}

Algorithm~\ref{alg-solution} is closely related to the well-known method of ``surface calculations" as described, for example, in \cite{UrBe:1986}.   One important difference, however, is that our version is decoupled from amplitudes, and makes clear that the method's validity is restricted to generic travel time vectors.  Illustrations of the algorithm's efficacy on synthetic data up to 16 layers can be found in the preprint \cite[Section~5.2]{Gi:Purely2012}.

\subsection{Local linearity of $\Phi$ on cells of $\ggg$\label{sec-local}}

Understanding the geometric structure of $\ggg$ makes it possible to give a precise description of the nonlinear nature of $\Phi$.   
The structure of the components (\ref{half-hyperplane}) of $\hhh_n$ implies that each $\ggg_n$ is a disjoint union of open convex sets, which will be referred to as cells, each of which is the positive cone over an open convex polytope. The details are as follows.   

For each $A$ arising as a solution to the factorization problem, let $\ccc_A$ denote the set of all $\tau\in\ggg$ such that $A\tau=\Phi(\tau)$.   Sets of the form $\ccc_A$ will be referred to as \emph{cells}.   Let $\wcc_A$ denote the subset consisting of normalized $\tau$,
\[
\wcc_A=\bigl\{\tau\in\ccc_A\,\bigl|\,\langle\mathbb{1},\tau\rangle=1\bigr\}.
\]
Because of (\ref{commutes}), it is evident that $\ccc_A=\real_{>0}\,\wcc_A$; in other words, $\ccc_A$ is the positive cone over $\wcc_A$.       

\begin{prop}\label{prop-convex}
Each normalized cell $\wcc_A\subset\real^{n+1}_{>0}$ is a convex polytope, open relative to the standard simplex 
\[
\Delta_{n}=\bigl\{x\in\real^{n+1}_{>0}\,\bigl|\,\langle\mathbb{1},x\rangle=1\bigr\}. 
\]
\end{prop}
\begin{pf}
Straightforward calculations show that $\ccc_A$ is convex and open relative to $\real^{n+1}_{>0}$, where $n$ is the number of columns of $A$.   Since $\wcc_A$ is the central projection of $\ccc_A$ onto $\Delta_{n}$, it follows that $\wcc_A$ is convex and open relative to $\Delta_{n}$.   Finally, the fact that the boundary of $\wcc_A$ is a finite union of sets of the form (\ref{half-hyperplane}) intersected with $\Delta_{n}$ implies that $\wcc_A$ is a convex polytope in $\Delta_{n}$. 
\end{pf}

The domain $\ggg$ is a disjoint union of cells $\ccc_A$; and each cell is the cone over a convex polytope, by Proposition~\ref{prop-convex}.  On each cell $\ccc_A$, the mapping $\Phi$ is linear, represented by the matrix $A$.   Thus,
\begin{thm}\label{thm-locally-linear}
The mapping $\Phi:\ggg\rightarrow\cup_{n=1}^\infty\real^{n+1}_{>0}$ is locally linear on each $\ggg_n$.   More precisely, for each $\tau\in\ccc_A\subset\ggg_n$, 
\[
\Phi(\tau)=A\tau.
\]
\end{thm}
This makes clear the nature of the mapping $\Phi$ on any given $\real^{n+1}_{>0}$.   As $\tau$ varies over any fixed cell $\ccc_A$ of $\ggg_n$, $\Phi$ is represented by the matrix $A$.  But when $\tau$ crosses the boundary of $\ccc_A$ and passes to another cell $\ccc_{A^\prime}$, then $\Phi$ switches to a different linear map $A^\prime$.   When $\tau$ crosses through a boundary face of the highest possible dimension, the representation (\ref{half-hyperplane}) of the boundary allows one to read off the relationship between $A$ and $A^\prime$:  either $A^\prime$ is obtained from $A$ by permuting two rows; or, in the case where $k^\prime=\mathbb{1}$, $A^\prime$ is obtained from $A$ by either adding a new bottom row or deleting the existing bottom row.   If $\tau$ crosses a lower-dimensional part of the boundary of $\ccc_A$, then several rows of $A$ may be permuted, or several new rows added etc.~to yield $A^\prime$.

\section{Localized amplitude inversion\label{sec-localized}}

If one is given $G^{(\tau,R)}(t)$ for $0\leq t\leq T$ in the form 
\[
\sum_{j=1}^da_j\delta(t-\sigma_j),
\]
and if one has determined $\tau$ using Algorithm~\ref{alg-solution}, then the amplitude polynomials can be correctly matched to the amplitude data $\{a_j\}$.  Explicitly, $a_j=a(R,k)$, where $k$ is the (generically unique) lattice point such that $\langle k,\tau\rangle=\sigma_j$.   How does one recover the reflection coefficients $R$ given the collection of formulas $a_j=a(R,k)$?   The easiest way is to isolate those reflection coefficients corresponding to primary lattice points $k^j$ consisting of 1's followed by 0's, 
\[
k^j_r=\left\{\begin{array}{cc}1&\mbox{ if }0\leq r\leq j\\
0&\mbox{ if }j<r
\end{array}\right..
\]
Then $R_0=a(R,k^0)$ and 
\begin{equation}\label{scheme}
R_{j+1}=\frac{a(R,k^{j+1})}{a(R,k^j)}R_j,
\end{equation}
whereby $R$ can be recovered recursively.   This is simple and fast, but it requires that each of the primary amplitudes $a(R,k^j)$ be accurate.   If any of these, say $a(R,k^j)$, is corrupted in the measured data, then the recursive scheme (\ref{scheme}) cannot recover $R_{j+r}$ for any $r\geq 0$.   In seismic data acquired on land, for  example, amplitudes associated with early arrival times are often badly distorted.  It is therefore of interest to find other ways to compute a given $R_j$ from the amplitude data that arrives beyond a certain time.

The three-term recurrence satisfied by Jacobi polynomials $P^{(\alpha,1)}_n$ gives a way to do this.   Our main result along these lines is what we call the \emph{eight amplitudes theorem}, which gives an explicit formula for the reflectivity $R_j$ in terms of a collection of amplitudes all of whose arrival times are greater than or equal to $\tau_0+\cdots+\tau_{j+1}$.   Several preparatory lemmas are needed to set the stage.

\subsection{Preparatory lemmas\label{sec-preparatory}}

The classical recurrence for $\alpha\geq-1$ and $N\geq0$ is:
\begin{multline}\label{recurrence}
(2N+4)(N+\alpha+3)(2N+\alpha+3)P^{(\alpha,1)}_{N+2}(z)=\\
(2N+\alpha+4)\left\{(2N+\alpha+5)(2N+\alpha+3)z+\alpha^2-1\right\}P^{(\alpha,1)}_{N+1}(z)\\
-2(N+\alpha+1)(N+2)(2N+\alpha+5)P^{(\alpha,1)}_N(z).
\end{multline}   
(See \cite[\S 4.5]{Sz:1975}.)  In conjunction with Theorem~\ref{thm-factor-Jacobi}, this recurrence provides the crucial ingredient for the eight amplitudes theorem:  a quadratic equation satisfied by the product $R_{j-1}R_{j+1}$.   We use the notation $e^j\in\integer^{n+1}$ for the $j$th standard basis vector (for which $e^j_i=\delta_{ij}$).    
\begin{lem}\label{lem-quadratic}
Fix integers $n, j, p, q$ such that: $n\geq2$, $1\leq j\leq n-1$, and min$\{p,q\}\geq 1$.   For $r=0,1,2$, let 
\[
k^r=(k^r_0,\ldots,k^r_n)\in\integer^{n+1}
\]
be any three lattice points satisfying the conditions
\[
(k^0_{j-1},k^0_j,k^0_{j+1})=(1,p,q),\quad k^1=k^0+e^j+e^{j+1},\quad k^2=k^0+2(e^j+e^{j+1}), 
\]
and, in the case  $j\leq n-2$, the additional condition $k^0_{j+2}=0$.   Let 
\[
x=(x_0,\ldots,x_n)\in[-1,1]^{n+1}
\]
and write 
\[
\tilde{a}_r=a(x,k^r)\quad (r=0,1,2).
\] 
Set $N={\rm min}\{p,q\}-1$ and $\alpha=|p-q|$, and define $Q_{p,q}$ to be the quadratic
\[
Q_{p,q}(y)=Ay^2+By+C,
\]
where 
\[
\begin{split}
A=&2(N+\alpha+1)(N+2)(2N+\alpha+5)\\
B=&(2N+\alpha+4)\left\{(2N+\alpha+5)(2N+\alpha+3)+\alpha^2-1\right\}\frac{\tilde{a}_1}{\tilde{a}_0}\\
C=&(2N+4)(N+\alpha+3)(2N+\alpha+3)\frac{\tilde{a}_2}{\tilde{a}_0}-\\
&2(2N+\alpha+4)(2N+\alpha+5)(2N+\alpha+3)\frac{\tilde{a}_1}{\tilde{a}_0}(x_{j-1}x_j^2x_{j+1})
\end{split}
\]
if $p\leq q$, and otherwise
\[
\begin{split}
A=&2(N+\alpha+1)(N+2)(2N+\alpha+5)\\
B=&(2N+\alpha+4)\left\{(2N+\alpha+5)(2N+\alpha+3)+\alpha^2-1\right\}\frac{p(q+1)}{q(p+1)}\frac{\tilde{a}_1}{\tilde{a}_0}\\
C=&(2N+4)(N+\alpha+3)(2N+\alpha+3)\frac{p(q+2)}{q(p+2)}\frac{\tilde{a}_2}{\tilde{a}_0}-\\
&2(2N+\alpha+4)(2N+\alpha+5)(2N+\alpha+3)\frac{p(q+1)}{q(p+1)}\frac{\tilde{a}_1}{\tilde{a}_0}(x_{j-1}x_j^2x_{j+1}).
\end{split}
\]
Then $Q_{p,q}(x_{j-1}x_{j+1})=0$, provided $\tilde{a}_0\neq0$.  
\end{lem}
\begin{pf}
If $p\leq q$, then by Theorem~\ref{thm-factor-Jacobi} and Definitions~\ref{defn-factor} and \ref{defn-amplitude},
\[
\begin{split}
\frac{\tilde{a}_1}{\tilde{a}_0}=&-x_{j-1}x_{j+1}\frac{P_{N+1}^{(\alpha,1)}(1-2x_j^2)}{P_N^{(\alpha,1)}(1-2x_j^2)}\\
\frac{\tilde{a}_2}{\tilde{a}_0}=&(x_{j-1}x_{j+1})^2\frac{P_{N+2}^{(\alpha,1)}(1-2x_j^2)}{P_N^{(\alpha,1)}(1-2x_j^2)};
\end{split}
\]
and if $p>q$ then 
\[
\begin{split}
\frac{p(q+1)}{q(p+1)}\frac{\tilde{a}_1}{\tilde{a}_0}=&-x_{j-1}x_{j+1}\frac{P_{N+1}^{(\alpha,1)}(1-2x_j^2)}{P_N^{(\alpha,1)}(1-2x_j^2)}\\
\frac{p(q+2)}{q(p+2)}\frac{\tilde{a}_2}{\tilde{a}_0}=&(x_{j-1}x_{j+1})^2\frac{P_{N+2}^{(\alpha,1)}(1-2x_j^2)}{P_N^{(\alpha,1)}(1-2x_j^2)}.
\end{split}
\]
In either case the equation $P_N^{(\alpha,1)}(1-2x_j^2)\;Q_{p,q}(x_{j-1}x_{j+1})=0$ reduces to the recurrence (\ref{recurrence}), multiplied by $(x_{j-1}x_{j+1})^2$ and with $z=1-2x_j^2$.   Therefore if $\tilde{a}_0\neq0$, and hence $P_N^{(\alpha,1)}(1-2x_j^2)\neq0$, it follows from the recurrence relation that $Q_{p,q}(x_{j-1} x_{j+1})=0$.   This proves the lemma.  
\end{pf}

We remark that one can obtain a similar result by replacing the condition $k^0_{j+2}=0$ with $k^0_{j+2}=1$, and making the obvious modifications.   Thus there is in fact a wider possible choice of triples $k^0,k^1,k^2$ to use in subsequent computations.  

The goal is to extract $x_j$ from the amplitude data.   The lemma shows that the amplitude data  $\tilde{a}_0,\tilde{a}_1,\tilde{a}_2$ together with the quantity $\xi=x_{j-1}x_j^2x_j$ determine the quadratic $Q_{p,q}$,  and that furthermore $x_{j-1}x_{j+1}$ is a root of $Q_{p,q}$.  Therefore, letting $y_+$ and $y_-$ denote the roots of $Q_{p,q}$, either 
\begin{equation}\label{either}
|x_j|=\sqrt{\xi/y_+}\quad\mbox{ or }\quad |x_j|=\sqrt{\xi/y_-}.
\end{equation}
The following two lemmas show that both the sign of $x_j$ and the quantity $\xi$ can also be expressed in terms of amplitude data.  In light of equation (\ref{either}), this suffices to determine $x_j$ (up to choosing the correct root of $Q_{p,q}$).   

\begin{lem}\label{lem-xi}
Fix integers $n, j, u, v$ such that: $n\geq2$, $1\leq j\leq n-1$, and min$\{u,v\}\geq 1$.  For $r=3,4,5,6$, let 
\[
k^r=(k^r_0,\ldots,k^r_n)\in\integer^{n+1}
\]
be any four lattice points satisfying the conditions
\[
\begin{split}
(k^3_{j-1},k^3_j,k^3_{j+1})=&(1,u,1),\quad k^4=k^3+e^j\\
(k^5_j,k^5_{j+1})=&(1,v),\quad k^6=k^5+e^{j+1},
\end{split}
\]
and, in case $j\leq n-2$, $k^5_{j+2}=0$.    Let 
\[
x=(x_0,\ldots,x_n)\in[-1,1]^{n+1}
\]
and write 
\[
\tilde{a}_r=a(x,k^r)\quad (r=3,4,5,6).
\] 
Then if $\tilde{a}_3 \tilde{a}_5\neq 0$, 
\[
x_{j-1}x_j^2x_{j+1}=\frac{u}{u+1}\frac{\tilde{a}_4 \tilde{a}_6}{\tilde{a}_3 \tilde{a}_5}.
\]
\end{lem}
\begin{pf}
This follows directly from the formulas in Definitions~\ref{defn-factor} and \ref{defn-amplitude}.  For instance, if $j\leq n-2$, 
\[
\begin{split}
\frac{\tilde{a}_4 \tilde{a}_6}{\tilde{a}_3 \tilde{a}_5}&=\frac{f^{(1,u+1)}(x_{j-1})f^{(u+1,1)}(x_j)f^{(1,v+1)}(x_j)f^{(v+1,0)}(x_{j+1})}{f^{(1,u)}(x_{j-1})f^{(u,1)}(x_j)f^{(1,v)}(x_j)f^{(v,0)}(x_{j+1})}\\
&=\frac{u+1}{u}x_{j-1}x_j^2x_{j+1}.
\end{split}
\]
The case $j=n-1$ is virtually the same.   
\end{pf}
\begin{lem}\label{lem-sign}
Fix integers $n, j, m$ such that: $n\geq2$, $1\leq j\leq n-1$, and $m\geq1$.  Let 
\[
k^7=(k^7_0,\ldots,k^7_n)\in\integer^{n+1}
\]
have the structure
\[
k^7=\mathbb{1}^j+2me^{j+1}.
\]
Let 
\[
x=(x_0,\ldots,x_n)\in[-1,1]^{n+1}
\]
and write $\tilde{a}_7=a(x,k^7)$.   Then if $\tilde{a}_7\neq0$,  $x_j/|x_j|=-\tilde{a}_7/|\tilde{a}_7|$.   
\end{lem}
\begin{pf}
According to Definitions~\ref{defn-factor} and \ref{defn-amplitude}, the given structure of $k^7$ implies that 
\[
\tilde{a}_7=(1-x_0^2)\cdots(1-x_{j}^2)(-x_j)^{2m-1}x_{j+1}^{2m}.
\]
Therefore $\tilde{a}_7$ and $x_j$ have opposite sign if $\tilde{a}_7\neq0$. 
\end{pf}

\subsection{Explicit formulas\label{sec-explicit}}

\begin{thm}[Eight amplitudes]\label{thm-eight-amplitudes}
Fix integers $n, j$ with $n\geq2$ and $j$ in the range $1\leq j\leq n-1$.   For $r=0,\ldots,7$, let $k^r\in\integer^{n+1}$ be any eight lattice points conforming to the hypotheses of Lemmas~\ref{lem-quadratic}, \ref{lem-xi} and \ref{lem-sign}.    Let 
\[
x=(x_0,\ldots,x_n)\in[-1,1]^{n+1}
\]
and write 
\[
\tilde{a}_r=a(x,k^r)\quad (r=0,\ldots,7).
\] 
Let $A,B,C$ be defined as in Lemma~\ref{lem-quadratic}, where in the formula defining $C$ the term $x_{j-1}x_j^2x_{j+1}$ is replaced with $u\tilde{a}_4 \tilde{a}_6/((u+1)\tilde{a}_3\tilde{a}_5)$,
so that $A,B$ and $C$ are completely determined by $\tilde{a}_0,\ldots,\tilde{a}_6$.   
If $\tilde{a}_0\tilde{a}_3\tilde{a}_5\tilde{a}_7\neq0$, then either 
\[
x_j=-\frac{\tilde{a}_7}{|\tilde{a}_7|}\sqrt{\frac{2uA\tilde{a}_4\tilde{a}_6}{(u+1)\tilde{a}_3\tilde{a}_5(-B+\sqrt{B^2-4AC})}},
\]
or
\[
x_j=-\frac{\tilde{a}_7}{|\tilde{a}_7|}\sqrt{\frac{2uA\tilde{a}_4\tilde{a}_6}{(u+1)\tilde{a}_3\tilde{a}_5(-B-\sqrt{B^2-4AC})}}.
\]
\end{thm}
\begin{pf}
Given Lemma~\ref{lem-quadratic}, the above formulas for $x_j$ result from combining Lemmas~\ref{lem-xi} and \ref{lem-sign} with formula (\ref{either}).  \end{pf}

Some remarks are in order regarding Theorem~\ref{thm-eight-amplitudes}.   Firstly, there are many different possible choices for the eight lattice points $k^0,\ldots,k^7$ arising in the hypotheses of Lemmas~\ref{lem-quadratic}, \ref{lem-xi} and \ref{lem-sign}, so there are many different ways to compute $x_j$.   In other words, the eight amplitudes formula is highly redundant.   Furthermore, this redundancy can be used to eliminate one of the two formulas, thereby determining $x_j$ uniquely.  Generically, if two different choices of $(k^0,k^1,k^2)$ are used to evaluate two different versions of the eight amplitudes formula for a fixed vector $x$, only the true value for $x_j$ will be common to the two versions.  The following corollary gives an explicit implementation of this.   A second remark concerning Theorem~\ref{thm-eight-amplitudes} is that the eight lattice points, and eight corresponding amplitudes, need not all be different---five lattice points suffice.  In fact, one can determine $x_j$ \emph{uniquely} using just seven amplitudes, as in the following example.  

\begin{cor}[Seven points formula]\label{cor-seven}
Fix integers $n, j$ with $n\geq2$ and $j$ in the range $1\leq j\leq n-1$.   For $r=0,\ldots,6$, define lattice points $k^r\in\integer^{n+1}$ as follows:
\[
\begin{array}{ccl}
k^0&=&\mathbb{1}^{j+1}\\
k^1&=&k^0+e^j+e^{j+1}\\
k^2&=&k^0+2e^j+2e^{j+1}\\
k^3&=&k^0+e^{j+1}\\
k^4&=&k^0+e^j\\
k^5&=&k^0+e^j+2e^{j+1}\\
k^6&=&k^0+2e^j+3e^{j+1}\\
\end{array}.
\]
Let 
\[
x=(x_0,\ldots,x_n)\in[-1,1]^{n+1}
\]
and write 
\[
\tilde{a}_r=a(x,k^r)\quad (r=0,\ldots,6).
\] 
Set 
\[
\begin{split}
y_{\pm}&=-\frac{7\tilde{a}_1}{5\tilde{a}_0}\pm\sqrt{3\frac{\tilde{a}_1\tilde{a}_3\tilde{a}_4}{\tilde{a}_0^3}+\left(\frac{7\tilde{a}_1}{5\tilde{a}_0}\right)^2-\frac{9\tilde{a}_2}{5\tilde{a}_0}}\\
Y_{\pm}&=-\frac{5\tilde{a}_5}{4\tilde{a}_3}\pm\sqrt{\frac{5\tilde{a}_4\tilde{a}_5}{2\tilde{a}_0^2}+\left(\frac{5\tilde{a}_5}{4\tilde{a}_3}\right)^2-\frac{4\tilde{a}_6}{3\tilde{a}_3}}.
\end{split}
\]
If $\tilde{a}_0\tilde{a}_3\neq0$ and $\frac{7\tilde{a}_1}{5\tilde{a}_0}\neq\frac{5\tilde{a}_5}{4\tilde{a}_3}$, then there is a unique $y$ such that 
\[
\{y\}=\{y_+,y_-\}\cap\{Y_+,Y_-\}.
\]
Furthermore,
\[
x_j=-\frac{\tilde{a}_3}{|\tilde{a}_3|}\sqrt{\frac{\tilde{a}_3\tilde{a}_4}{2\tilde{a}_0^2y}}.
\]
\end{cor}
\begin{pf}
The hypothesis of Lemma~\ref{lem-quadratic} is satisfied by the triple of lattice points $(\tilde{k}^0,\tilde{k}^1,\tilde{k}^2)$ for both
\[
(\tilde{k}^0,\tilde{k}^1,\tilde{k}^2)=(k^0,k^1,k^2)\quad\mbox{ and }\quad (\tilde{k}^0,\tilde{k}^1,\tilde{k}^2)=(k^3,k^5,k^6).  
\]
The first case corresponds to $(p,q)=(1,1)$ and the second to $(p,q)=(1,2)$.  Also, the hypothesis of Lemma~\ref{lem-xi} is satisfied by 
\[
(\tilde{k}^3,\tilde{k}^4,\tilde{k}^5,\tilde{k}^6)=(k^0,k^4,k^0,k^3)
\]
(with $(u,v)=(2,2)$), whereby $x_{j-1}x_j^2x_{j+1}=\frac{a_3a_4}{2a_0^2}$.  Using this, Lemma~\ref{lem-quadratic} gives $y_\pm$ as the roots of $Q_{1,1}$ and $Y_\pm$ as the roots of $Q_{1,2}$.  Since $\tilde{k}^7=k^3$ satisfies the hypothesis of Lemma~\ref{lem-sign} (with $m=1$), it follows by Theorem~\ref{thm-eight-amplitudes} that 
\[
x_j=-\frac{\tilde{a}_3}{|\tilde{a}_3|}\sqrt{\frac{\tilde{a}_3\tilde{a}_4}{2\tilde{a}_0^2y}},
\]
for some $y\in\{y_+,y_-\}$ and some $y\in\{Y_+,Y_-\}$, provided $a_0a_3\neq0$.  The condition 
\[
\frac{7\tilde{a}_1}{5\tilde{a}_0}\neq\frac{5\tilde{a}_5}{4\tilde{a}_3},
\]
guarantees that $Q_{1,1}$ and $Q_{1,2}$ do not have both roots in common, so there is at most one possible choice of $y$.   This completes the proof.  
\end{pf}

The fact that the seven points formula of Corollary~\ref{cor-seven} is homogeneous (of degree 0) in the amplitudes $\tilde{a}_0,\ldots,\tilde{a}_6$ has practical significance.   If the measured data represented by the reflection Green's function $G^{(\tau,R)}(t)$ is distorted by an unknown scalar multiple (so that one is given $\alpha\, G^{(\tau,R)}(t)$ for $0<t<T$, where $\alpha$ is unknown),  then the calculation of $R_j$ using the seven points formula is unaffected.    This means that reflection coefficients can in principle be recovered accurately even from data that is recorded using a miscalibrated instrument.   The same observation applies to Theorem~\ref{thm-eight-amplitudes}.   Although it is less immediate from the theorem's statement, the formulas for $x_j$ are homogeneous in the amplitudes and therefore invariant under a uniform rescaling.

\section{Conclusion}

From the mathematical perspective, the results in Sections~\ref{sec-amplitudes} and \ref{sec-polydisk} illuminate some surprising connections between the PDEs governing wave propagation in layered media on one hand and Jacobi polynomials, holomorphic functions on the polydisk, and almost periodic functions on the other hand.   
Beyond their intrinsic mathematical interest, these connections are also shown in Sections~\ref{sec-power}, \ref{sec-arrival} and \ref{sec-localized} to have practical implications for the inverse theory relevant to imaging modalities.   

For instance, the result of Section~\ref{sec-power} relates to deconvolution techniques in geophysical imaging, as follows.  An experimentally generated plane wave acoustic source is not perfectly impulsive and so is usually modeled by a compactly supported source wavelet $f(t)$, representing the short duration motion of a piezoelectric disk, a vibroseis plate, or some other physical mechanism.   The measured reflection data is then properly expressed as a convolution of the form $f\ast G^{(\tau,R)}$, from which one wants to extract $G^{(\tau,R)}$.  This entails a straightforward deconvolution if $f$ is known, but in certain contexts $f$ is not known, necessitating a strategy for blind deconvolution such as the following.   Assuming that 
\[
|\widehat{G^{(\tau,R)}}(\omega)|\cong 1,
\]
 the power spectrum of the measured data $f\ast G^{(\tau,R)}$ is simply the power spectrum of the source wavelet $f$.   Assuming further that $f$ is minimum phase (see \cite{GiLa:2012}), it can be recovered from its power spectrum, in turn allowing $G^{(\tau,R)}$ to be extracted from the measured data.   Theorem~\ref{thm-unity} supports the first of these two assumptions, thereby providing mathematical justification for a longstanding geophysical supposition.   However, there is a second geophysical assumption that is less well justified, historically made necessary by the absence of an explicit representation for $G^{(\tau,R)}$.   It is assumed in \cite[Chapter~2]{BlCoSt:2001} (among other sources) that the reflectivity sequence approximates $G^{(\tau,R)}$, 
\[
\sum_{j=0}^nR_j\delta\bigl(t-(\tau_0+\cdots+\tau_j)\bigr)\cong G^{(\tau,R)}(t).
\]
But direct numerical comparison does not substantiate this assumption at all (see \cite[Section~4]{Gi:Comb2013}).   On the contrary, one has to extract the reflectivity sequence, or equivalently $(\tau,R)$, from the reflection Green's function.  Algorithm~\ref{alg-solution} and Corollary~\ref{cor-seven} provide a direct method for doing so.     

Section~\ref{sec-arrival} describes some key geometric features of the mapping  
\[
(\tau,R)\mapsto G^{(\tau,R)},
\]
revealing a contrast between local and global behaviour.  Arrival times generically decouple from amplitudes, and, locally, arrival times depend linearly on layer depths while amplitudes depend algebraically on reflection coefficients.  But this local characterization does not extend globally, since a different linear-algebraic relationship holds for different cells in the $\tau$ domain, as described in Section~\ref{sec-local}.   The decoupling of travel time inversion means that it can be carried out first, and then used to determine the correct correspondence between amplitude data and amplitude polynomials necessary for the analysis of amplitudes in Section~\ref{sec-localized}.  

As far as we know, Theorem~\ref{thm-eight-amplitudes} and its accompanying Corollary~\ref{cor-seven}   constitute the first proof that it is possible in principle to recover reflection coefficients using amplitude data from a restricted time window, or, in other words, to carry out localized amplitude inversion.  More than this,  the formulas of Theorem~\ref{thm-eight-amplitudes} and Corollary~\ref{cor-seven} determine a particular reflection coefficient directly, without requiring prior computation of any other reflection coefficients.  From the perspective of imaging, this has the important implication that computational effort can be concentrated on zones of interest.

\bibliography{ReferencesEAL}

\begin{thebibliography}{10}

\bibitem{Be:1955}
A.~S. Besicovitch.
\newblock {\em Almost periodic functions}.
\newblock Dover Publications Inc., New York, 1955.

\bibitem{BlCoSt:2001}
N.~Bleistein, J.~K. Cohen, and J.~W. Stockwell, Jr.
\newblock {\em Mathematics of multidimensional seismic imaging, migration, and
  inversion}, volume~13 of {\em Interdisciplinary Applied Mathematics}.
\newblock Springer-Verlag, New York, 2001.
\newblock Geophysics and Planetary Sciences.

\bibitem{BrGo:1990}
L.~M. Brekhovskikh and O.~A. Godin.
\newblock {\em Acoustics of Layered Media {I}}, volume~5 of {\em Springer
  Series on Wave Phenomena}.
\newblock Springer, Heidelberg, 1990.

\bibitem{FoGaPaSo:2007}
J.-P. Fouque, J.~Garnier, G.~Papanicolaou, and K.~S{\o}lna.
\newblock {\em Wave propagation and time reversal in randomly layered media},
  volume~56 of {\em Stochastic Modelling and Applied Probability}.
\newblock Springer, New York, 2007.

\bibitem{Gi:Purely2012}
P.~C. Gibson.
\newblock The purely singular 1-{D} acoustic reflection problem.
\newblock 47 pages, arXiv:1206.2695 [math-ph], June 2012.

\bibitem{Gi:Comb2013}
P.~C. Gibson.
\newblock The combinatorics of scattering in layered media.
\newblock 24 pages, arXiv:1305.3961 [math.CO], May 2013.

\bibitem{GiLa:2012}
P.~C. Gibson and M.~P. Lamoureux.
\newblock Identification of minimum-phase-preserving operators on the
  half-line.
\newblock {\em Inverse Problems}, 28(6):065020, 13, 2012.

\bibitem{Mi:2006}
J.~Milnor.
\newblock {\em Dynamics in one complex variable}, volume 160 of {\em Annals of
  Mathematics Studies}.
\newblock Princeton University Press, Princeton, NJ, third edition, 2006.

\bibitem{MoNePa:2008}
R.~V. Moody, M.~Nesterenko, and J.~Patera.
\newblock Computing with almost periodic functions.
\newblock {\em Acta Crystallogr. Sect. A}, 64(6):654--669, 2008.

\bibitem{Oh:2002}
T.~Ohsawa.
\newblock {\em Analysis of several complex variables}, volume 211 of {\em
  Translations of Mathematical Monographs}.
\newblock American Mathematical Society, Providence, RI, 2002.
\newblock Translated from the Japanese by Shu Gilbert Nakamura, Iwanami Series
  in Modern Mathematics.

\bibitem{Ru:1969}
W.~Rudin.
\newblock {\em Function theory in polydiscs}.
\newblock W. A. Benjamin, Inc., New York-Amsterdam, 1969.

\bibitem{Sz:1918}
G.~Szeg{\H{o}}.
\newblock Ein {B}eitrag zur {T}heorie der {P}olynome von {L}aguerre und
  {J}acobi.
\newblock {\em Math. Z.}, 1(4):341--356, 1918.

\bibitem{Sz:1975}
G.~Szeg{\H{o}}.
\newblock {\em Orthogonal polynomials}.
\newblock American Mathematical Society, Providence, R.I., fourth edition,
  1975.
\newblock American Mathematical Society, Colloquium Publications, Vol. XXIII.

\bibitem{UrBe:1986}
B.~Ursin and K.-A. Berteussen.
\newblock Comparison of some inverse methods for wave propagation in layered
  media.
\newblock {\em Proceedings of the IEEE}, 74(3):389--400, 1986.

\end{thebibliography}

\end{document}